\documentclass[12pt]{article}
\title{Two-stage Optimization with Recourse and Revocation}
\author{Haotian Jiang}
\usepackage{amsmath,amsfonts,latexsym,graphicx}
\usepackage{fullpage,color}
\usepackage{bm}
\usepackage{multicol}
\usepackage{url,hyperref}
\usepackage[linesnumbered,boxed,ruled,vlined]{algorithm2e}

\newtheorem{lemma}{Lemma}
\newtheorem{theorem}[lemma]{Theorem}

\newtheorem{model}{Model}

        {\medskip}

        {\hspace*{\fill}$\Box$\par}
        {\hspace*{\fill}$\Box$\par\vspace{4mm}}
        {\hspace*{\fill}$\Box$\par}

\newcommand{\etal}{{\em et al.}\ }

\newcommand{\E}{\textbf{E}}

\newcommand{\opt}{\textrm{\sc OPT}}

\begin{document}
\maketitle

\begin{abstract}
Two-stage optimization with recourse model is an important and widely used model, which has been studied extensively these years. In this article, we will look at a new variant of it, called the two-stage optimization with recourse and revocation model. This new model differs from the traditional one in that one is allowed to revoke some of his earlier decisions and withdraw part of the earlier costs, which is not unlikely in many real applications, and is therefore considered to be more realistic under many situations. We will adopt several approaches to study this model. In fact, we will develop an LP rounding scheme for some cover problems and show that they can be solved using this scheme and an adaptation of the rounding approach for the deterministic counterpart, provided the polynomial scenario assumption. Stochastic uncapacitated facility location problem will also be studied to show that the approximation algorithm that worked for the two-stage with recourse model worked for this model as well. In addition, we will use other methods to study the model.
\end{abstract}

\section{Introduction}

\subsection{Two-stage optimization with recourse}
Uncertainty is an inevitable facet of many decision problems in real world. And stochastic optimization, by incorporating uncertainty into the input data, is a leading approach to model such uncertainty. The study of stochastic optimization problem stems from the work of Dantzig \cite{129-4} and Beale \cite{129-1} in the 1950s, as is pointed out by  \cite{129}, and has received considerable attention from the community of theoretical computer science. Among those problems that were studied intensively were the so-called two-stage optimization with recourse problem, which model several infrastructure planning problems in which decisions would have to be made under some degree of uncertainty about the requirements. Although more accurate decisions could be made when the actual requirement is revealed, the cost of this recourse action is inevitably inflated until then.

As an illastrating example, consider a company planning to build some facilities to serve the public. In the first stage, the company only know the distribution of the clients, and is able to build some facilities with cost $f_i$. Later when the set of clients materialize, the company has to built some more facilities so that the cost of connecting each client to at least one of the opened facilities is not too high. However, the recourse action would be much more expensive than the first stage action. Particularly, there is a $\lambda>1$ denoting the ratio between the costs in the two stages, called the inflating factor. In general, $\lambda$ might be different for each facility. The problem is: given the distribution of clients, the cost of connecting clients to facilities and the cost of building facilities in both stages, how should we make decisions such that the total expected cost is minimized. This is the famous SUFL (stochastic uncapacitated facility location) problem which we will mention in later sections.

There have been several methods to solve the two-stage optimization with recourse problem. As is introduced in \cite{main}, an LP rounding approach is used by Swamy \etal in  \cite{125}. We also know of a method that reveal the connection between the stochastic version and the deterministic vertion of the problem known as boosted sampling, which was invented by Gupta \etal  \cite{80}.

\subsection{Our model}
Despite that the above mentioned model captures several important aspects of reality, it nevertheless can be improved to be more realistic and in this paper, we propose a variation of it. In many applications, it is possible that companies start off buying something and later on regret their actions by selling these objects for lower prices. We model such sellings as a revocation reaction and in this way we obtain a new model in which we are allowed to revoke some of the choices we made in the first stage. Such model is called a two-stage optimization with recourse and revocation model. It is worth mentioning that this model is in fact a natural generalization of the previous model. By setting the profit gained from revoking first-stage action to 0, we degerate to the previous model. Based on our knowledge at present, we know of no studies conducted on such a model so this article is in fact a pioneering attemp to solve problems that fall into the category of this model. Here we will give an example of our model. Although we could formulate a high-level abstract discription as was done in  \cite{80}, we choose to be specific and propose a concrete example.
\begin{model}[Rooted Steiner tree]
We are given an edge-weighted graph G(V,E) with edge weight $\omega_{e}$ for each edge e satisfying triangular inequality and a root $r\in V$. We are also given a distribution of the possible clients. In the first stage, we only know the distribution and can buy a set of edges $F_0$ whereas in the second stage, once the real set of clients is revealed, we can buy some more edges $F_2$ for $\lambda\omega_e$ and sell some useless edges $F_1\subseteq F_0$ for $(1-\sigma)\omega_e$ to obtain a feasible steiner tree that connect all clients. The model asks to minimize $\omega(F_0)-(1-\sigma)E[\omega(F_1)]+\lambda E[\omega(F_2)]$.
\end{model}
Notice that the above model is equivalent to the following model, which is called two-stage optimization with recourse and reservation:
\begin{model}[Rooted Steiner tree]
We are given an edge-weighted graph G(V,E) with edge weight $\omega_{e}$ for each edge e satisfying triangular inequality and a root $r\in V$. We are also given a distribution of the possible clients. In the first stage, we only know the distribution and can book a set of edges $F_0$ for $\sigma\omega_e$ whereas in the second stage, once the real set of clients is revealed, we can buy some more edges $F_2$ for $\lambda\omega_e$ and buy some edges $F_1\subseteq F_0$ reserved in the first stage for $(1-\sigma)\omega_e$ to obtain a feasible steiner tree that connect all clients. The model asks to minimize $\sigma\omega(F_0)+(1-\sigma)E[\omega(F_1)]+\lambda E[\omega(F_2)]$.
\end{model}
In this article, we will only work on the second model which is easier to handle. And also from now on, we will use the names of these two models interchangeably since they are equivalent. Moreover, all problems studied in the following sections are required to satisfy sub-additivity, the formal definition of which can be found in  \cite{80}. Informally, it means that the union of solutions to instances is a solution to the union of these instances.

In section 2, we will develope a linear programming rounding technique to tackle some problems. And in section 3, we will improved on the approximation algorithms for the problems mentioned in section 2. And in section 4, we will turn to other methods.
\subsection{Our result}
\begin{enumerate}
\item We will develope a LP rounding scheme for the cover problems and show that the deterministic LP rounding methods can be adapted to this model using the scheme. We will show a $2k$-approximation algorithm for the vertex cover problem and a $O($log $n) k$-approximation algorithm for the set cover problem, where $k$ is $(\lambda+\sigma-1)/(2-2\sigma)$, which is not bad if $\sigma$ is not too large. Moreover, we will develope a double randomized rounding technique for handling our model if a randomized rounding technique exists for the deterministic problem.
\item The SUFL (Stochastic Uncapacated Facility Location) problem will be shown to be solvable by a rounding approach of the recourse model  \cite{125-22}, which in fact bears great resemblance to the rounding method for the deterministic problem. This gives a 5-approximation algorithm.
\item We will improve on the result of $1$ and $2$ by showing that the covering problems can be achieved the same approximation factor as their deterministic counterparts, by generalizing the ideas of \cite{127}. We will also show an alternative 3.81-approximation algorithm for the SUFL problem generalizing the ideas of the Shmoys-Swamy \cite{125} algorithm for the two-stage with recourse model and the algorithm of Swamy's for the deterministic problem \cite{125-27}.
\item We will show a trivial method that simply ignores the ability to revoke decisions. It turns out that this method works pretty well if $\sigma$ is very large. Also, we will give a heuristic algorithm which we believe will work. However, a bound on the performance is required.
\end{enumerate}

\section{LP Rounding Technique, A First Attempt}
In this section, LP rounding approach will be adopted to yield approximation algorithms for our model. In all but the last subsection, it will be assumed that the number of scenarios is polynomial in the input, (known as polynomial scenario assumption), which enable us to write out explicitly the LP and solve it efficiently. We will begin by a generalized rounding scheme and use it to solve several cover problems. Then we will turn to the SUFL problem. All the above problem will be handled under the assumption that the number of distribution of scenarios is polynomial in the input. In the last part of this section, we will refer to a theorem proved by  \cite{31} using a generalized method, which demonstrates that the more general case of a black box model can be degerated to the polynomial scenario model, loosing a factor of only $(1+O(\epsilon))$.
\subsection{A generalized LP rounding approach}
We illustrate this rounding scheme by using the set cover LP as an example.
\begin{model}[2-stage set cover with recourse and revocation]
We are given a ground set of elements $F$, a family of subsets $S$ with cost $\omega_s$ for subset $s\in S$ and a distribution $\pi$ of the scenarios. We assume polynomial scenario in which the support of the distribution is polynomial in the input. In the first stage, we can reserve a set of subsets $F_0$ paying a cost of $\sigma \omega_s$ for each $s$. In the second stage, the real set of clients is revealed, we can buy some more edges $F_2$ for $\lambda\omega_e$ and pay for some reserved edges $F_1\subseteq F_0$ for $(1-\sigma)\omega_e$ such that $F_1\cup F_2$ covers all the clients. We always assume that $0<\alpha<1$ and $\lambda>1$. The model asks to minimize $\sigma\omega(F_0)+(1-\sigma)E[\omega(F_1)]+\lambda E[\omega(F_2)]$.
\end{model}

We use an LP-rounding approach to solve the problem.
\begin{eqnarray}
&Minimize:&\sigma\sum_{s\in S}x_s \omega_s+(1-\sigma)\sum_{A,s\in S}P_A \omega_s y_{A,s}+\lambda\sum_{A,s\in S}P_A \omega_s z_{A,s}\\
&\forall A,e\in A&:\sum_{s:e\in s}(y_{A,s}+z_{A,s})\geq1\\
&\forall A,e\in A&:y_{A,s}\leq x_{s}\\
&\forall A,e\in A&:x_{s},y_{A,s},z_{A,s}\geq 0
\end{eqnarray}

The difficulty in adopting the rounding algorithm of the deterministic set cover is that the two stages in our model is hard to seperate. There is an additional link between the two stages due to the reservation (or revocation) action. Therefore, we have to first process our solution before we we can use LP rounding. Also, there's yet another difficulty: the integer solution we obtained from rounding should still satisfy that $y_{A,s}\leq x_s$. This is the same as asking: if we are given a problem together with a smaller subproblem problem whose the solution is smaller than the original problem, can we round both solutions so that the rounded solution to the subproblem is still smaller than the rounded solution to the original problem? We show this can be done for the cover problems.

We do a first-round processing of the LP solution $(x_s^*,y_{A,s}^*,z_{A,s}^*)$ to $(x_s,y_{A,s},z_{A,s})$ s.t. $\forall A,e\in A, \sum_{s:e\in s}y_{A,s}$ is either always no less than 1/2 or is always less than 1/2 (Of course we can change this factor but it turns out that this doesn't improve on the performance guanrantee). To do this, we consider each e, if the above is desobeyed, we decrease all $y$s to a half (one $y$ can only be decreased once, if it has been decreased before, we turn to other $y$s), and increase the $z$s correspondingly. Notice that this step pays a factor within $k=(\lambda+\sigma-1)/(2-2\sigma)$. This is not bad given $\sigma$ is not too large. If $\sigma$ is not far from 1, then intuitively, we can simply buy all subsets we reserved earlier. Therefore, we can first process all the $y$s to be equal to their corresponding $x$s, this will increase the value of the solution to at most $\frac{1}{\sigma}$ and then we can use whatever rounding technique for the two-stage optimization with recourse to solve our problem. In fact, we will show in the next section that this later method of buying all the reserved subsets is quite general, and can be used together with any approximation algorithm for the two-stage recourse model, provided $\sigma$ is not too small.

\subsection{The set cover problem}
As is known, there is a randomized rounding technology  \cite{Randomized Rounding} that finds a solution in at most O($\alpha$)OPT with high probability, where $\alpha=$ln $n$. Now we show that this randomized rounding technique can be adopted to our problem, by a scheme we call double randomized rounding.

Suppose that we first solve the stochastic LP and process the solution using the above approach. The rounding of the solution $(x_s,y_{A,s},z_{A,s})$ to $(x'_s,y'_{A,s},z'_{A,s})$must satisfies that $x'_s$ denotes the set of subsets reserved in the first stage. And $y'_{A,s}\leq x_s,z'_{A,s}$ denotes the subsets that are bought in the second stage. The rounding approach is as follows: Let E=\{e:$\sum_{s:e\in s}y_{A,s}\geq 1/2 (\forall A)$\}. Notice that for any $e\in E,\sum_{s:e\in s}x_s\geq 1/2$. So \{$2x_s$\} is a fractional solution to E, we can round it to a solution with cost at most $O(\alpha)\sum_s 2x_s \omega_s$. Specifically, the randomized rounding is as follows: in each round, we choose a set $s$ with probability $x_s$ and we do this over and over until all elements in $E$ is covered. It is not hard to show that the number of rounds is at most $O(\alpha)$. Also in each round, we record the set of subsets chosen for future use. Now when the scenario A is revealed, $\{2y_{A,s}\}$ constitute a solution to elements in E, and $\{2z_{A,s}\}$ constitute a solution to the elements in $A\backslash E$. We can apply the same randomized rounding technique for elements in $A\backslash E$, but for elements in E, we have to combine the results together with that when rounding all $x$s to guarantee that $y_{A,s}\leq x_s$ still holds after the rounding. In each round, we pick $S$ with probability $y_{A,s}/x_s$ and then take the intersection of all the subsets chosen with the corresponding stage in the first stage. By the same analysis, we can obtain a solution within $O(\alpha)$ times the optimal with high probability. In fact, the factor hidden behind the $O$ is very small: the $O(\ln n)$ is actually $\ln n+O(\ln\ln n)$.

Therefore, we can obtain a solution with cost at most $O(\alpha)((1-\sigma)\sum_s 2y_{A,s}+\lambda\Sigma_s 2z_{A,s})$. Therefore, we obtain a $2kO(\alpha)-$approximation algorithm for the above model.

It is worth pointing out that this double rounding scheme is quite general. For any problem which can be solved by a randomized rounding algorithm, it is well likely that using double randomized rounding, we can solve its two-stage with recourse and revocation counterpart. This idea will be further developed in later sections.
\subsection{The vertex cover problem}
\begin{model}[2-stage vertex cover with recourse and revocation]
The model is almost the same as the set cover. In the vertex cover, we are given a graph with vertex set $S$ and edge set $E$. The vertex set we choose from $S$ is required to cover the clients in $E$ in the sense that each client is adjacent to some vertex we choose. The remaining is exactly the same.
\end{model}

We begin by writing out the stochastic LP for the vertex cover problem, which is almost the same as that for the set cover problem.
\begin{eqnarray*}
&Minimize:&\sigma\Sigma_{s\in S}x_s \omega_s+(1-\sigma)\sum_{A,s\in S}P_A \omega_s y_{A,s}+\lambda\Sigma_{A,s\in S}P_A \omega_s z_{A,s}\\
&\forall A,e\in A&:\sum_{s:s\in e}(y_{A,s}+z_{A,s})\geq1\\
&\forall A,e\in A&:y_{A,s}\leq x_{s}\\
&\forall A,e\in A&:x_{s},y_{A,s},z_{A,s}\geq 0
\end{eqnarray*}

As is known, independently rounding all those no less than 1/2 to 1 and those less than 1/2 to 0 yields a 2-approximation algorithm for the deterministic model. The algorithm for this model is along the lines in the last subsection. By the same token, we can get a 4k-approximation algorithm for this model.

\subsection{Uncapacitated facility location}
\begin{model}[2-stage SUFL with recourse and revocation]
In the 2-stage SUFL with recourse and revocation problem, we are given a set $F$ of facility locations, a set $C$ of possible clients, and a distribution on the scenario. Here, we assume the scenarios are given explicitly as an input, as well as the demand $d_j^k$ of each client $j$. Each facility has a ground cost $f_i^0$ and a scenario cost $f_i^k$. The distance between each client $j$ and each facility $i$ is given by $c_{ij}$, which satisifies the triangular inequality. In the first stage, we are allowed to reserve a set of facilities with $\sigma$ times the ground cost. In the second stage, when the real demand of each client materializes, we are allowed to pay for a subset of reserved facilities for $(1-\sigma)$ times the ground cost, and also recourse several other facilities for the scenario cost. Then we have to assign each client to an opened facility. It is always assumed that $0<\sigma<1$ and that the scenario cost of each facility is higher than the ground cost.The goal is to minimize the expected total cost for opening facilities plus the expected total cost of connecting all the clients in each scenario.
\end{model}
The relaxation of the linear program of SUFL with recourse and revocation is :
\begin{eqnarray*}
&\min:&\qquad \sigma\Sigma_{i\in F}f_i^0 y_i^0+\sum_{k=1}^{m}p_k((1-\sigma)\sum_{i\in F}f_i^0y_i^k+\\
&&\qquad \sum_{i\in F}f_i^kz_i^k+\sum_{i\in F,j\in D}d_j^kc_{ij}x_{ij}^k)\\
&s.t.:&\\
&\forall j,k&\qquad \sum_{i\in F}x_{ij}^k\geq d_j^k\\
&\forall i,k&\qquad     y_i^k\leq y_i^0\\
&\forall i,j,k& \qquad   x_{ij}^k\leq y_i^k+z_i^k\\
&\forall i,j,k&\qquad x_{ij}^k,y_i^0,y_i^k,z_i^k\geq 0
\end{eqnarray*}
In the following argument, we assume that demand $d_j^k$ only takes on value $0$ or $1$, yet this constraint can easily be removed, since irrational demand can be simulated by rational demand to any degree of precision and rational demand can be simulated by 0-1 demand by multiplying a least common multiple and then cut each client into several parts. It is worthy to point out that although the above LP fixed the connection cost, the idea can be generalized to the case when the connection cost depends on the scenaro and that the demand take other values as well. In fact, we can directly absorb the connection cost factor into the demand factor. Also it is noticable that here we don't require uniform inflating factor.

Notice the idea adopted by  \cite{125-22} can be applied to this model as well. Their idea generalized the approximation algorithm proposed in  \cite{125-26}. We now show how to use their method to obtain an approximation algorithm for our model. Suppose we solve the relaxation of the LP to obtain an optimal fractional solution: $(x,y,z)$. We begin by adopting a filtering technique used in  \cite{125-22-22}. We fix a constant $0<\alpha<1$ which will be specified later. For each client-scenario pair $(j,k)$, define the fractional service cost, as was done in  \cite{125-22}, to be $c_{jk}^*=\sum_i c_{ij}x_{ij}^k$. Now we define a "neighborhood distance" of a client-scenario pair $(j,k)$ to be the smallest distance $c_{jk}^{\alpha}$ such that: $\sum_{i:c_{ij}\leq c_{ij}^{\alpha}}x_{ij}^k\geq \alpha$. This guarantees that facilities in this neighborhood satisfies a fraction of at least $\alpha$ times the total need. And it is easy to prove that: $c_{jk}^\alpha \leq \frac{1}{1-\alpha}c_{jk}^*$. If on the contrary that this doesn't hold true, then we consider all those facilities of distance greater than $c_{jk}^\alpha$ and we would discover that the contribution of these facilities to the client exceeds $c_{jk}^*$, which is a contradiction. Now we filter our solution to a new solution with the following property: each client-scenario pair is assigned to only facilities within its neighborhood. Denote the new solution as $(\overline{x},\overline{y},\overline{z})$. Set:
$$
\overline{x_{ij}^k}=\left\{
\begin{array}{ccl}
\min\{1,\frac{1}{\alpha}x_{ij}^k\} &          &c_{ij} \leq c_{jk}^\alpha\\
 0                                           &          &   otherwise
\end{array}\right.
$$
And set $\overline{y_i^k}=\min\{1,\frac{1}{\alpha}y_{i}^k\}$ and $\overline{z_i^k}=\min\{1,\frac{1}{\alpha}z_{i}^k\}$. It is intuitively clear that this is a feasible solution.We fix another variable $0<\beta<1$ which will be specified later.

Then we round as follows: initially, we set every client "unserved" and define $F^0$ and $F^k$ to be empty. Now we consider $(j,k)$ with the smallest fractional service cost. Define $S^0$ to be the set of facilities satisfying: $\overline{x_{ij}^k}>0\wedge \overline{y_{i}^k}>0$. Similarly,  Define $S^k$ to be the set of facilities satisfying: $\overline{x_{ij}^k}>0\wedge \overline{z_{i}^k}>0$.

If $\sum_{i\in S^0}\overline{y_i^k}\geq \beta$, we are in a first-stage dominant situation. Thus we seek a facility $i\in S^0$ such that $f_i^0$ is the smallest. We open $i$ and add it to $F^0$, by setting $y_i^0=1, y_i^k=1$. For all other facilities $i'\in S^0\cup S^k$, set $y_{i'}^0=y_{i'}^k=0$. For each client-scenario pair $(j',k')$such that there exists a facility  $i'\in S^0\cup S^k$ with $c_{i'j'}\leq c_{j'k'}^\alpha$, set $x_{i,j'}^{k'}=1$ and mark them "served". By doing this, we are actually satisfying all those client-scenario pairs whose neighborhood overlap with $(j,k)$. And if on the other hand that $\sum_{i\in S^0}\overline{y_i^k}<\beta$, it must be that $\sum_{i\in S^k}\overline{z_i^k}\geq \beta$. This time we are in a recourse-stage dominant situation. Similarly, we choose a facility $i$ in $S^k$ with smallest $f_i^k$. We add $i$ to $F^k$ by setting $z_i^k=1$ and then close all other facilities in the neighborhood of $(j,k)$. For clients $(j',k)$ (Notice we require the scenario to be the same this time) whose neighborhood overlaps that of $(j,k)$'s, we assign them to facility $i$ and mark them "served".

We recurrently follow the above step until all client-scenario pairs are eventually assigned to some facility. Then $F^0$ is the set of facilities to be opened in the first stage and $F^k$ is the set of facilities to be recoursed in the second stage. We can easily show that the above algorithm works in our model. Since in the above process, each facility is touched at most once in the whole procedure, we can guarantee that we do not make contradictory decisions. Moreover, whenever we choose to set a $y_i^k$ to be $1$, we also set the corresponding $y_i^0$ to be one. So our integral solution is feasible. Now we are ready to show something similar to Lemma $1$ in  \cite{125-22}. In fact, it is merely a generalization to it.
\begin{lemma}The solution $(x,y,z)$ we get by the rounding algorithm satisfies:
(1) For each client-scenario pair $(j,k)$, if $x_{ij}^k=1$, then $c_{ij}\leq 3\frac{1}{1-\alpha}c_{jk}^*$. (2) $\sum_{i\in F}f_i^0y_i^0\leq\frac{1}{\beta}\sum_{i\in F}f_i^0\overline{y_i^0}$. (3) For each scenario k: $\sum_{i\in F}f_i^k y_i^k\leq \frac{1}{\beta}\sum_{i\in F}f_i^k\overline{y_i^k}$. (4) For each scenario k: $\sum_{i\in F}f_i^k z_i^k\leq \frac{1}{1-\beta}\sum_{i\in F}f_i^k\overline{z_i^k}$.
\end{lemma}
The proof for the above lemma is straightforward and would be omitted here. It follows that the above algorithm has an approximation factor of $\max\{\frac{3}{1-\alpha}, \frac{1}{\alpha\beta}, \frac{1}{\alpha(1-\beta)}\}$. By setting $\beta=1/2$ and $\alpha=2/5$, we obtain a $5$-approximation algorithm for the SUFL with recourse and revocation model:
\begin{theorem}
Under the polynomial scenario assumption, there is a 5-approximation algorithm that runs in polynomial time for the 2-stage stochastic uncapacitated facility location with recourse and revocation problem.
\end{theorem}
If the reservation ratio $\sigma$ is very big, we can again simply first let each $y_i^k=y_i^0$ which raises the cost of the solution by a factor of at most $\frac{1}{\sigma}-1$. Then we use the algorithm by Shmoys \etal in  \cite{125} to obtain a approximation factor of $\frac{3.225}{\sigma}$. Notice this is better than the former algorithm for large $\sigma$. In the next section, we will present an algorithm that achieves better performance. However, the algorithm presented in this section is also of particular interest because it is a simple adaptation of an the ideas used for the deterministic problem \cite{125-26}.\\

\subsection{The black-box model}
As  \cite{main} mentioned, the black box model is actually a more general model in which we are not given the distribution but only a procedure through which we can generate samples. Notice that the total number of scenarios may be exponential and therefore renders it impossible to solve the stochastic LP in polynomial time. Nevertheless,  \cite{125} \cite{130} \cite{31} show that if the maximum recourse ratio $\lambda$ is polynomially bounded, it suffice to draw only a polynomial number of samples from the black box. Specifically, there is a polynomial approximation scheme for the black box model.  \cite{129} uses an adaptation of the ellipsoid algorithm to handle the black box model, and  \cite{130} showed, by using subgradient, that we only need polynomial number of samples under some mild condition. Later Charikar \etal  \cite{31} provided a more general method: they show that either by repeating SAA (sample average approximation) many times or rejecting high-cost scenarios, we can obtain a solution within $(1+O(\epsilon))\opt$ with high probability. They considered a general form of stochastic program:
\begin{equation}
\min_{x\in X}f(x)=c(x)+\E_{\omega}[q(x,\omega)]
\end{equation}
And the SAA method draws $N$ samples:
\begin{equation}
\hat{f}(x)=c(x)+\frac{1}{N}\sum_{i=1}^Nq(x,\omega_i)
\end{equation}
In fact, they show the following theorem:
\begin{theorem}[Repeating SAA]
Consider a collection of k functions $\hat{f}^1, \cdots, \hat{f}^k$, such that $k=\Theta(\epsilon^{-1}log \delta^{-1})$ and the $\hat{f}^i$ are independent sample average approximations of the function f, using $N=\Theta(\lambda^2\epsilon^{-4}\cdot k\cdot log |X| log \delta^{-1})$ samples each. For $i=1,\cdots, k$, let $\overline{x}^i$ be an $\alpha-approximate$ minimizer of the function $\hat{f}^i$. Let $i=\arg\min_j \hat{f}^j(\overline{x}^j)$. Then, with probability $1-3\delta$, $\overline{x}^i$ is an $(1+O(\epsilon))\alpha-$approximation minimizer of $f(\cdot)$.
\end{theorem}
There is yet another theorem adopting the methods of rejecting high-cost scenarios. This method is more efficient in that it draws way fewer samples. However, the above theorem suffices for our purpose. Notice that the stochastic LP of our model lies in the category of their general program and therefore, we can turn the black box model into a polynomial scenario model with high probability by losing only a multiplicative factor of $(1+\epsilon)$. If, in addition, that our approximation algorithm for the polynomial scenario model is a constant factor one, the additional $\epsilon$ factor due to the black-box model is also an additive one.\\
\section{Improved approximation algorithms}
The algorithms in the previous section are generally nonsatisfying. Particularly, in our rounding algorithm for the cover problems, the $\lambda$ on the numerator of the factor $k$ can yield poor performance of the algorithm when $\lambda$ is very large, which is often the case in many applications. The approximation factor for the SUFL problem is also far behind the approximation factor for the same problem under the two-stage with recourse model, which achieves an approximation factor of 2.369 \cite{127}. In this section, we illustrate some improved algorithms for the same problems under the polynomial scenario model.
\subsection{Cover problems revisited}
In this section, we generalize the idea used in \cite{127} to develope approximation algorithms for our model. Srinivasan \cite{127} took a key view of the multi-stage recourse problem as an online process and then use randomized rounding to solve it. He assumed that the optimal fractional solution arrived in an online fashion: in each stage, only the value of the variables for that stage is revealed, and we are to make irrevocable decisions. It seems at first glance that this problem is much harder than the original one in which we know all the information of the solution. However, it was showed surprisingly that we are able to do as better as the best algorithm for the deterministic problem. In this subsection, we show that Srinivasan's idea can be generalized to handle our model as well, yielding better approximation factor than algorithms in the last section.
\subsubsection{Set cover revisited}
The relaxation of the linear programming is the same as that in the last section and suppose we obtained an optimal fractional solution $(x_s,y_{A,s},z_{A,s})$. We choose a parameter $\lambda$ which will be specified later. Let $x'_s=\min\{\lambda x_s,1\}$, $y'_{A,s}=\min\{\lambda y_{A,s},1\}$ and $z'_{A,s}=\min\{\lambda z_{A,s},1\}$. In the first stage, reserve each subset with probability $x'_s$. In the second stage, choose each reserved subset with probability $y'_{A,s}/x'_s$, and recourse every subset with probability $z'_{A,s}$ (then take the union with those reserved and paid for). Suppose the total number of elements is $n$, then we take $\lambda=\ln n+\psi(n)$, where $\psi(n)$ is function that grows slowly of $n$ and that $\lim_{n\rightarrow \infty} \psi(n)=\infty$. It is easy to observe that the expected cost is at most $\lambda$ times that of the opimal fractional solution. Also by the same analysis as in the last section, for any element $i$, the probability that $i$ is not covered is at most $\exp(-\lambda)=\exp(-\psi(n))/n$. Thus using the union bound, the total probability of failure is at most $\exp(-\psi(n))$. This achieves an approximation factor of $(1+o(1))\ln n$.

\begin{theorem}
Under the polynomial scenario assumption, there is a $(1+o(1))\ln n$-approximation algorithm that runs in polynomial time for the 2-stage stochastic set cover with recourse and revocation problem.
\end{theorem}
\subsubsection{Vertex cover revisited}
In fact, as is pointed out in \cite{127}, the vertex cover is just a special case of a general kind of set cover: each element appears in at most a constant $b$ number of subsets. In the vertex cover problem, $b=2$, and it is easy to show that with minor changes, this method works for other values of $b$ as well.

Now assume that we have obtained the optimal fractional solution $(x,y,z)$ to the stochastic LP. We first multiply the solution by 2, i.e. $x'_s=\min\{2x_s,1\}$ and similarly for $y$ and $z$. In the first stage, we reserve each vertex with probability $x'_s$ and in the second stage, we first buy each reserved vertex with probability $y'_{A,s}/x'_s$. For each vertex in the realized scenario $A$, reserved or not, if it is not bought by $y'_{A,s}$ and $y'_{A,s}+z'_{A,s}\geq 1$, then we recourse it using $z'_{A,s}=1$(If this happens to be a reserved vertex, then we simply set the corresponding $y'_{A,s}$ to be 1 instead of $z'_{A,s}$, which will not increase the cost). If $y'_{A,s}+z'_{A,s}<1$, then we know that this vertex is not perchased by $(x',y',z')$ so we simply set $z'_{A,s}$ to be 0.

The analysis for the above algorithm is fairly standard. First we notice that if a vertex is bought in the solution $(x',y',z')$, then it is bought by the algorithm. Next we bound the expected cost of the solution obtained: the first stage expected cost is essentially the same as that of $(x',y',z')$. For the second stage cost, notice that paying for reserved verices cost exactly the same as $(x',y',z')$ in expectation. For the recourse actions, if a vertex is not perchased by $(x',y',z')$, then its recourse stage cost is 0 and must be less than that of the solution $(x',y',z')$. Else, the chances that we recourse a particular vertex is at most $1-y'_{A,s}$ which is no greater than $z'_{A,s}$. This shows that the expected cost is bounded by the cost of $(x',y',z')$, which is bounded by 2 times the optimal fractional solution.

\begin{theorem}
Under the polynomial scenario assumption, there is a 2-approximation algorithm that runs in polynomial time for the 2-stage stochastic vertex cover with recourse and revocation problem.
\end{theorem}

\subsection{A better approximation algorithm for the SUFL problem}
In this section, we provide a better algorithm to the SUFL with recourse and revocation problem. This algorithm is a generalization of Shmoys and Swamy's algorithm for the two-stage with recouse problem \cite{125} and Swamy's algorithm for the deterministic UFL problem \cite{125-27}. In fact, Swamy's algorithm is a variant on a former algorithm due to Chudak and Shmoys \cite{125-27-18}. The Chudak and Shmoys algorithm is of particular interest to us because they use a randomized rounding technique in their algorithm. We develope the ideas of all the above algorithms to solve SUFL under our model. In the following sections, we first introduce the Chudak-Shmoys algorithm (CS algorithm for short) and then show the variant introduced by Swamy. For the purpose of conciseness, we only sketch their algorithms, and interested reader are refer to the reference for the details of the proof. Finally, we show how their ideas, when combined with the approximation algorithm by Shmoys and Swamy \cite{125}, can be adopted to yield an approximation algorithm for our model.
\subsubsection{The CS algorithm for UFL}
The formal definition for the deterministic problem can be found in \cite{125-27}. The CS algorithm makes use of the dual solution as well as the primal solution. We now write out explicitly the primal program:
\begin{eqnarray*}
&\min& \sum_if_iy_i+\sum_j d_j\Sigma_ic_{ij}x_{ij}\\
&\forall j&\sum_ix_{ij}	\geq 1\\
&\forall i, j&x_{ij}\leq y_i\\
&\forall i, j&x_{ij},y_i\geq 0
\end{eqnarray*}
And the dual program:
\begin{eqnarray*}
&\max& \sum_i\alpha_i\\
&\forall i, j& \alpha_j\leq d_j c_{ij}+\beta_{ij}\\
&\forall i& \sum_j\beta_{ij}\leq f_{i}\\
&\forall i, j& \alpha_{j},\beta_{ij}\geq 0
\end{eqnarray*}
The weak duality states that the objective value of any solution to the dual program cannot exceeds that of any solution to the primal problem. And strong duality states that the optimal solution of the dual program and the optimal solution of the primal program gives the same objective value. The CS algorithm assumes we solve the two programs and return an optimal primal solution $(x,y)$ and an optimal dual solution $(\alpha,\beta)$. This can be easily done under the polynomial scenario assumption. They use a definition first introduced by Shmoys, Tardos \& Aardal \cite{125-26} known as $g-$close solution: A solution $(x,y)$ is called $g-$close if for every $j$, $x_{ij}>0\Rightarrow c_{ij}\leq g_j$. Chudak and Shmoys notice that the solution to the primal solution is in fact $\alpha-$close due to complementary slackness. They then use a randomized rounding technique to select which facility to open. It is worth mentioning that their algorithm for any $g-$close solution, and not just for the optimal solution to the LP programs. We now sketch their algorithm below.

For convenience, we assume that for any $i$ and $j$, if $x_{ij}>0$, then $x_{ij}=y_i$. This is called a complete solution and in fact, they show that any solution can be reduced to a complete solution by deviding the facilities into several fractional parts. The CS algorithm first pick out some non-intersecting clusters and assign each client that is not in any cluster to a near cluster as its representative. There is exactly one facility opened in any cluster and facilities disjoint with any cluster are opened at random. Let $C_j=\sum_ic_{ij}x_{ij}$ be the cost incurred by the LP solution to assign client $j$, and denote as $F_j$ all the facilities client $j$ is fractionally assigned to. And the algorithm works as follows:\\
\textbf{(1)}. Order all the clients in increasing $C_j+\alpha_j$ value (denote this list by $S$), and repeatedly do the following until $S$ becomes empty: choose the client with the smallest value and form a cluster around it with all facilities in $F_j$. Remove from $S$ the every client (including  $j$) that is served by some facility in $F_j$ and make $j$ the representative of each such client. Let $D$ be the set of cluster centers.\\
\textbf{(2)} Within each cluster $F_j, j\in D$, open exactly one facility by choosing facility $i$ with probability $y_i$.\\
\textbf{(3)} Each non-central facility is opened independently with probability $y_i$.\\
\textbf{(4)} We assign each client to the nearest open facility.

They are able to show the following:
\begin{lemma}
For any client $j$, we have $\E[X_j]\leq C_j+\frac{2}{e}\alpha_j$.
\end{lemma}
And since $\sum_i \alpha_i$ is no larger than the optimal solution, it immediately follows that the algorithm is a $(1+\frac{2}{e})$-approximation algorithm.
\subsubsection{Swamy's algorithm}
Swamy generalized the CS algorithm by first performing a filtering. His algorithm is more general in that it does not require any knowledge of the actual demand of the clients. This property is very important and this will become clear later. Swamy's algorithm takes in a complete solution $(x,y)$. Again, if the solution is not complete, Swamy shows that he can first process the solution to make it complete without changing the objective value. For convenience, we will assume that the solution is complete. Swamy uses parameters $0<\gamma<1$ and $r=1/\gamma$ and his algorithm proceeds as follows.

Sort the facilities in $F_j$ in increasing $c_{ij}$ value. Let $i'$ be the first facility such that $\sum_{i\leq i'}x_{ij}\geq \gamma$. Define $R_j(\gamma)=c_{í'j}$ and $C_{ij}(\gamma)=(\sum_{i<i}ç_{ij}x_{ij}+c_{i'j}(\gamma-\sum_{i<i'}x_{ij}))/\gamma$. Let $N_j\subseteq F_j$ be the facilities up to including $i'$ in the sorted order.

Again, for simplicity, assume that each $y_i\leq \gamma$ and for any client $j$, $\sum_{i\in N_j}y_i$ is exactly equal to $\gamma$. Swamy shows that this assumption can be easily removed. Under such assumption, $C_j(\gamma)=(\sum_{i\in N_j}c_{ij}x_{ij})/\gamma$. Now we get a new solution $\hat{x_{ij}}=x_{ij}/\gamma$ if $i\in N_j$ and 0 otherwise and $\hat{y_{ij}}=y_i/\gamma$. Then simply run the CS algorithm on this new solution.

Swamy did a more refined analysis on the algorithm and was able to show the following lemma:
\begin{lemma}
For any $\gamma\geq 1/3$, and any client $j$, $\E[X_j]\leq (1+e^{-r}\cdot \frac{1+\gamma}{1-\gamma})C_j$.
\end{lemma}
And by setting $\gamma=\frac{1}{1.7}$, we get an 1.705-approximation algorithm, which is better than the CS algorithm and also acquired a demand-oblivious property. Now we are ready to show an approximation algorithm for the SUFL problem.

\subsubsection{Approximation algorithm for SUFL under our model}
The linear program is the same as that of section 2.4 and we solve it to obtain an optimal solution $(x,y,z)$. As is done by Shmoys and Swamy, we can always split the assigned value for each client: $x_{ij}^k=x_{ij}^{k1}+x_{ij}^{k2}$ such that $x_{ij}^{k1}\leq y_i^k$ and that $x_{ij}^{k2}\leq z_i^k$. We fix a parameter $\theta=2.29/(2.29+1.52)$ for simplicity. Notice that each client-scenario pair $(j,k)$ is either served by $x_{ij}^{k1}$ to no less than $\theta$ or is served by $x_{ij}^{k2}$ to no less than $1-\theta$. The first situation means that $\sum x_{ij}^{k1}\geq \theta$ and the second situation means that $\sum x_{ij}^{k2}\geq 1-\theta$. We now try to use our solution to create a deterministic problem to decide which facility to open in the first stage and which to pay for in the second stage. To do so, we consider the client-scenario pair as before. For each client-scenario pair $(j,k)$ to satisfy the first condition (call these \emph{first-stage client-scenario pair} and otherwise \emph{second-stage client-scenario pair}), we place it with demand $p_k$. And for those second-stage client-scenario pair, we do not consider it at this point, because we are going to simply use the solution to the recourse stage to satisfy their needs. In their algorithm, Shmoys and Swamy was able to first produce a fractional assigment that is independent of any scenario, and can therefore merge each client-scenario pair of a certain client into one \cite{125}. But this can not be done in our model since we are not able to obtain such a scenario-independent fractional assignment. But instead, we will turn to Swamy's algorithm introduced earlier.

\textbf{(1)} We first deal with first-stage client-scenario pair and see each client-scenario pair as an \emph{independent client}, although many of them may represent the same client. We first get a feasible solution $(\hat{x},\hat{y})$ by multiplying our solution with $\frac{1}{\theta}$. To be formal, $\hat{x}_{ij}^{k1}=\min\{1,x_{ij}^{k1}/\theta\}$ and $\hat{y}_i^k=\min\{1,y_i^k/\theta\}$. Then process the solution as is done in Swamy's algorithm and we use the same parameter $\gamma$ and $r=1/\gamma$. We assume the solution we get from this processing is $(\overline{x},\overline{y})$, where $\overline{x}_{ij}^{k1}=\min\{1,r\hat{x}_{ij}^{k1}\}$ for "near" facilities and 0 for "far" facilities (the formal discription of this can be found in section 2.5.2), and  $\overline{y}_{i}^{k}=\min\{1,r\hat{y}_{i}^{k}\}$. After this processing, we begin to run the CS algorithm on the new solution $(\overline{x},\overline{y})$ by first finding all the clusters and the representatives for all clients not in any cluster. Within each cluster, unlike the CS algorithm, we choose to reserve \emph{at least} one facility to open by picking each facility with probablity equal to $\overline{y}_i^0$. The requirement that at least one facility is reserved can be guaranteed by choosing each facility with probablity $\overline{y}_i^0$ and if no facility is chosen, then repeat this process. Then for those facility not in any cluster, we reserve each one with probablity $\overline{y}_i^0$. In the second stage, when the set of clients materialize, we pay for each facility reserved in the first stage $i$ with probablity $\overline{y}_i^k/\overline{y}_i^0$ if it is not in any cluster. For all those facilities in a cluster $F_j$, we pick each reserved facility with probablity $\overline{y}_i^k/\overline{y}_i^0$ in a dependent fashion to ensure that \emph{exactly one} facility is opened in each cluster.

\textbf{(2)} For those second-stage client-scenario pair, since $(\min\{1,\frac{1}{(1-\theta)}z\},\min\{1,\frac{1}{1-\theta}x\})$ is a solution in the recourse stage and we know the demand exactly, we can use the 1.52-approximation algorithm in \cite{125-20} to solve the recourse stage problem. Now after all these are done, we simply assign each client to its nearest facility.

\subsubsection{Analysis}
Now we analyze the performance of the algorithm. To make our notation more simple, we use $Cost(x)$ to denote the cost in the objective function due to $x$. For instance, $Cost(y^k)$ is simply the cost of the optimal LP solution due to paying for reserved facilities. First consider the first stage action and the second-stage action of buying reserved facilities (notice these are the actions to deal with first-stage client-scenario pair). To analyze these actions, we have to specify one process: the process of reserving facilities within each cluster. Because the sum of the probability to open each facility in a cluster is no less than 1, it follows that in each round, the probability that no facility is opened is at most $1/e$. If this event happens, we have to do another round. We therefore can bound the expectation of the cost of this process by: $\frac{1}{1-1/e}Cost(\overline{y}^0)$, which is bounded by $\frac{e}{1-e}\frac{r}{\theta}Cost(y^0)$. We let $\eta=\frac{e}{1-e}$, so the above can be writen as $\frac{r\eta}{\theta}Cost(y^0)$. The expectation cost of the action of buying reserved facilities can be bounded similarly: it cannot exceeds $\eta Cost(\overline{y}^k)$, which is bounded by $\frac{r\eta}{\theta}Cost(y^k)$. As is show by Swamy's anlysis \cite{125-27}, the expected assigment cost of first-stage client-scenario pair is raised by at most $(1+e^{-r}\cdot \frac{1+\gamma}{1-\gamma})/\theta=(1+e^{-r}\cdot \frac{r+1}{r-1})/\theta$. And finally, the cost of the recourse stage cost for opening facilities is bounded by $1.52/(1-\theta)$ times the corresponding cost in optimal solution to the LP. And combing these observations, the approximation factor of this algorithm is $\max\{\frac{r\eta}{\theta},\frac{r\eta}{\theta},(1+e^{-r}\cdot \frac{r+1}{r-1})/\theta,1.52/(1-\theta)\}$. By setting $r=1.447$, we get an 3.81-approximation algorithm for the two-stage SUFL with recourse and revocation problem, which is better than the algorithm in section 3.4 except when $\sigma$ is nearly 1 where we can simply buy every facility we reserved. We conclude this analysis into the following theorem:
\begin{theorem}
Under the polynomial scenario assumption, there is a 3.81-approximation algorithm that runs in polynomial time for the 2-stage stochastic uncapacitated facility location with recourse and revocation problem.
\end{theorem}



\section{Other methods, the Stochastic Steiner Tree problem}
In this section, we will explore other methods to solve our problem. We illustrate everything by the rooted steiner tree model in section 1. But it should be noticed that these methods can be generalized to other problems as well.
\subsection{A general method}
There is a very intuitive method to handle our model: by simply ignoring the fact that we are allowed to revoke earlier dicisions and in this subsection, we look at this idea. Now we only work on the two-stage optimization with recourse and reservation model. Notice that if $\sigma$ is not too small, the above idea gives a trivial $\frac{\beta}{\sigma}$-approximation method. We can simply solve the problem using boosted sampling, or whatever $\beta-$approximation algorithms there are for the normal two stage optimization problem $(1,\lambda)$, where the first number indicates the first stage cost ratio and $\lambda$ is the inflated factor in the second stage, without revoking any edges (which is equivalent to buying all reserved edges), which returns a result at least $\frac{\beta}{\sigma}$ times the optimal value for the normal 2-stage problem $(\sigma,\lambda)$, but this is smaller than \opt, since ignoring the action of actually buying the reserved edges in the second stage can obtain a solution to the $(\sigma,\lambda)$ problem. Spcifically, for the 2-stage stochastic rooted steiner tree problem, the best algorithm for the recourse problem is a 3.55-approximation algorithm, even under the black box model. Therefore, we get a $\frac{3.55}{\sigma}-$approximation algorithm for the rooted steiner tree problem under our model.

But if $\sigma$ is small, this algorithm yields terrible result. Intuitively, buying all the reserved edges is the lease we want to do if buying them is cost nearly as much as their original costs. Notice that if $\sigma$ is small, it is less expensive to make reservations and more expensive to pay for them in the second stage so intuitively, we should make more reservations yet pay only for those we needed after the clients materialize. And we are seeking an algorithm in which we make a large number of reservations and a practical way to make use of these reservations to obtain a practically cheap result for the real requirements once revealed. This intuition give rise to the following heuristic algorithm.

\subsection{Sampling heuristic}
In boosted sampling methods by  \cite{80}, they augment a solution by zeroing out all the edges and run Prim's algorithm on it. An intuitive way to make augmentation in our model is instead of zeroing out the reserved edges, we decrease their weights $\omega_e$ to $(1-\sigma)\omega_e$ and solve the deterministic problem using the approximation algorithm for the deterministic problem. Moreover, since we want to make more reservations in the case of small $\sigma$, we are to seek the number of samplings at least proportional to $\frac{\lambda}{\sigma}$. In the following, we suppose that we have an $\alpha-$approximation algorithm for the deterministic problem. The heuristic algorithm is as follows.

\begin{algorithm}[h]
\caption{Sampling Heuristic}
\label{sampling heuristic}
Sample $\lceil\frac{\lambda}{\sigma}\rceil$ samples $D_1, \cdots, D_{\lceil\frac{\lambda}{\sigma}\rceil}$ and take thir union $D$\;
Apply the $\alpha-$approximation algorithm to solve $D$ and obtain the first-stage edges\;
When the real requirement $S$ is revealed, solve the problem using the $\alpha-$approximation algorithm by changing the costs of edges in $D$ to $(1-\sigma)$ the normal scale and the edges not in $D$ to $\lambda$ the normal cost\;
{\bf Return} the total cost;
\end{algorithm}

W.l.o.g. we can assume that $\frac{\lambda}{\sigma}$ is a integer, so that we can get rid of the ceiling. We denote the sets of edges reserved in the first stage, those paid for in the second stage and those recoursed in the second stage by the optimal solution as $(F_0^*,F_1^*,F_2^*)$, and the optimal value as $Z^*=\sigma F_0^*+\E[(1-\sigma)F_1^*+\lambda F_2^*]$ (For notational simplicity, we use $F^*$ to denote both the set of edges as well as the total ground cost of these edges). Denote the set of clients materializes by $S$. Correspondingly, we use $(F_0,F_1,F_2)$ to denote the sets in the solution returned by our algorithm. Set $m=\frac{\lambda}{\sigma}$. We now bound the performance of our algorithm.

We first derive a bound for the first-stage reservation cost. We now assume that instead of satisfying the set of clients $S$, we want the optimal solution to satisfy the set of clients $D_i$, given $F_0^*$. This will return a second stage solution $(F_1^i,F_2^i)$. It is straightforward that $(F_1^i,F_2^i)$ is the same as $(F_1^*,F_2^*)$ in distribution since both $S$ and $D_i$ are samples drawn from the scenario distribution. Suppose that $F_D^1=\cup_iF_1^i$ and $F_D^2=\cup_iF_2^i$. It follows that $F_D^1$ is a subset of $F_0^*$. Notice that $F_D^1\cup F_D^2$ is a feasible soluion for the set of clients $D$. And therefore, by definition, our first stage cost $\sigma F_0\leq\alpha \sigma \E(F_D^1+F_D^2)\leq\alpha (\sigma F_0^*+\sigma m\E(F_2^*))\leq \alpha Z^*$.

Now we bound the second stage cost. Let $\xi(K,j)$ be the cost-sharing function, where we assign each client $j$ in every set of clients $K$ a value equals to $1/2$ times the cost of connecting it to its parent in a minimum spanning tree found by Prim's algorithm on that set of clients. For simplicity of notation, if $J$ is a set of clients, denote $\xi(K,J)$ as the sum of the cost shares of all clients in J, when the spanning tree is computed on $K$. We are now ready to develop a bound on the second stagge cost as follows: $\E[(1-\sigma)F_1+\lambda F_2]\leq 2(1-\sigma)\xi(S\cup D,D)+2\lambda \xi(S\cup D,S\backslash D)$. But $2(1-\sigma)\xi(S\cup D,D)\leq \frac{2(2-\sigma)}{\sigma}Z^*$ and $+2\lambda \xi(S\cup D,S\backslash D)\leq \frac{2\lambda}{\sigma m}Z^*=2Z^*$. Therefore, the second stage cost is bounded by $(2\frac{1-\sigma}{\sigma}+2)Z^*$.

Since there is a 1.39-approximation algorithm for the rooted steiner tree problem, $\alpha=1.39$. And by the above analysis, this is a $(3.39+\frac{2(1-\sigma)}{\sigma})-$approximation algorithm, which was slightly better for small $\sigma$.


\section{Concluding Remarks}
In the above sections, we briefly studied some of the problems in our new model. Yet the performance of these zre generally poorer than the performance of the best existing algorithms for the two-stage with recourse model, although we have shown that many ideas used to solve the recouse model can be generalized to solve our problem as well. Yet we do not, in general, know whether algorithms that match the performance of the best algorithms for the recouse model can be developed in the context of our model.

In addition, all problems in this article require that the revocation-ratio is uniform, which might be far from reality. Typically, in many applications, some actions are more irrevocable than others. A more general approach to handle non-uniform revocation-ratio is required.

Nevertheless, we write this paper to evoke further study on problems falling in the category of this model and variants of it that take more realistic considerations.
\section{Acknowledgements}
I would like to thank my tutor Prof. Jian Li for his help and suggestions in my doing this research, and for his survey paper  \cite{main} which gives rise to my idea of this new model.
\bibliographystyle{plain}

\end{document}